# Phyllosilicate Emission from Protoplanetary Disks: Is the Indirect Detection of Extrasolar Water Possible?

Melissa A. Morris[1, 2] and Steven J. Desch[1]

Keywords: water, protoplanetary disks, infrared emission, phyllosilicates

ABSTRACT

Phyllosilicates are hydrous minerals formed by interaction between rock and liquid water and are commonly found in meteorites originating in the asteroid belt. Collisions between asteroids contribute to zodiacal dust, which therefore reasonably could include phyllosilicates. Collisions between planetesimals in protoplanetary disks may also produce dust containing phyllosilicates. These minerals possess characteristic emission features in the mid-infrared and could be detectable in extrasolar protoplanetary disks. Here we determine whether phyllosilicates in protoplanetary disks are detectable in the infrared using instruments such as those on board the *Spitzer Space Telescope* and *SOFIA* (Stratospheric Observatory for Infrared Astronomy). We calculate opacities for the phyllosilicates most common in meteorites and compute the emission of radiation from a protoplanetary disk using a 2-layer radiative transfer model. We find that phyllosilicates present at the 3% level lead to observationally significant differences in disk spectra, and should therefore be detectable using infrared observations and spectral modeling. Detection of phyllosilicates in a protoplanetary disk would be diagnostic of liquid water in planetesimals in that disk and would demonstrate similarity to our own Solar System. We also discuss use of phyllosilicate emission to test the "waterworlds" hypothesis, which proposes that liquid water in planetesimals should correlate with the inventory of short-lived radionuclides in planetary systems, especially $^{26}$Al.

Running Title: Phyllosilicate Emission
Correspondence should be directed to:
Melissa A. Morris, Arizona State University, School of Earth and Space Exploration, PO Box 871404, Tempe, AZ 85287-1404
Telephone: 480-727-6221, Fax: 480-965-8102

[1]Arizona State University
[2]Missouri State University

## INTRODUCTION

*Terrestrial Planets and Water*

In the search for extraterrestrial life, it is extremely important to attempt to detect water in extrasolar planetary systems, particularly on terrestrial planets. The central requirements for life as we know it are a source of free energy, a source of carbon, and liquid water (Chyba *et al.* 2000). As Chyba *et al.* (2000) put it, "where there is liquid water, there is the possibility of life as we know it". So far the detection of water on a terrestrial exoplanet has not been achieved, although there are hints that they should have liquid water. Close-in extrasolar giant planets (hot Jupiters) have been observed to have water vapor in their atmospheres (Beaulieu *et al.* 2008) and water vapor emission from a protoplanetary disk was recently observed as well (Watson *et al.* 2007; Carr &

Najita 2008). However, the search for liquid water on *Earth-like* planets is extremely difficult. Should an Earthlike planet be discovered, strategies exist for detecting liquid water on the surface (Williams & Gaidos 2008). As of late 2008, though, no such planets have been found; the most Earthlike planet yet discovered is Gliese 581c (Udry *et al.* 2007, Selsis *et al.* 2007), which is nearly 5 times the mass of Earth.

Another approach to the search for liquid water on terrestrial planets is to follow the path upstream to the water's source. While not universally accepted, it is generally thought that the majority of Earth's water was delivered by planetesimals from the outer asteroid belt (Morbidelli *et al.*, 2000; Raymond *et al.*, 2004, Mottl *et al.* 2007). We refer the reader to these papers, whose arguments we attempt to summarize here. Comets have long been suggested as a source of Earth's water (Owen & Bar-Nun 1995), but the low terrestrial D/H ratio suggests otherwise (Drake & Righter, 2002). In fact, in the few comets for which the D/H ratio has been measured, it is twice that for Earth; the D/H ratio would not be reduced by chemical fractionation processes (Eberhardt *et al.* 1995, Bockelee-Morvan *et al.* 1998, Meier *et al.* 1998). In addition, comets are predicted to introduce too much Ar and other noble gases to be consistent with the low terrestrial Ar/$H_2$O ratio (Swindle & Kring 1997; Owen & Bar-Nun 1995; Morbidelli *et al.* 2000). Finally, the likelihood of sufficient comets colliding with the Earth is too low to account for the Earth's volatile content (Levison *et al.*

2001, Morbidelli *et al.* 2000). Drake & Righter (2002) claim that during its formation, Earth received no more than 50% of its water from comets; based on dynamical arguments, Morbidelli *et al.* (2000) claim the fraction is less than 10% of the Earth's present water budget.

Both Drake & Righter (2002) and Morbidelli *et al.* (2000) instead strongly argue that the Earth accreted "wet", with asteroids or planetesimals from the outer asteroid belt as the main source of the water. The D/H ratio in VSMOW (Vienna standard mean ocean water) is consistent with that of carbonaceous chondrites, both having D/H ratios of 150 ppm (Drake & Righter, 2002). As reviewed by Morbidelli *et al.* (2000), carbonaceous chondrites, associated spectrally with C-type asteroids (Gradie & Tedesco 1982) and believed to have formed in the outer asteroid belt (i.e., beyond 2.5 AU), contain ~ 10 wt% water (structurally bound in clays), whereas ordinary and enstatite chondrites, associated spectrally with S- and E-type asteroids from the inner belt (Gradie & Tedesco 1982), contain ~ 0.5-0.1 wt% water. Accretion of a few percent of Earth's mass from carbonaceous chondritic material from beyond 2.5 AU is sufficient to explain Earth's volatile content (Morbidelli *et al.* 2000; Mottl *et al.* 2007). Complications to this hypothesis include the difference in oxygen isotopic content between Earth and carbonaceous chondrites, and the abundances of siderophiles like Os carried by carbonaceous chondrites (Drake & Righter 2002). However, reasonable refutations to these objections exist (Mottl *et al.* 2007). We therefore proceed under the assumption that Earth and extrasolar terrestrial planets acquired their water during accretion, from planetesimals. Because of this, the volatile content of the planetesimals themselves will largely govern how much water or other volatiles a planet will eventually possess. Of course volatiles may be lost during impacts, and the fraction of water that should be sequestered in the mantle vs. outgassed to the surface is not known. All other things being equal, if the planetesimals that make up a terrestrial planet have twice the amount of water than the planetesimals that made up Earth, we may reasonably expect that planet to have twice as much water in its oceans than does Earth.

*Planetesimal Volatiles and $^{26}$Al.*

Desch & Leshin (2004) pointed out that if the volatile content of terrestrial planets is determined by the volatile content of the planetesimals from which it formed, then ultimately it will depend on the abundance of $^{26}$Al in the planetary system. A general consensus exists that the internal heating of asteroids in the Solar System was due to the presence of radioactive $^{26}$Al ($t_{1/2}$ = 0.7 Myr) in these bodies (Grimm & McSween 1993, hereafter GM93; Lugmair & Shukolyukov 1993; Huss *et al.* 2001; McSween *et al.* 2002, Gilmour & Middleton, 2009). GM93 explain the heliocentric zoning of the asteroid belt based on this heat source. At the time of the formation of calcium-rich, aluminum-rich inclusions (CAIs), the $^{26}$Al abundance in the Solar System was $^{26}$Al/$^{27}$Al = 5 x 10$^{-5}$ (MacPherson *et al.* 1995). Asteroids that reach sizes capable of trapping radiogenic heat (roughly 30 km in diameter) in the first 2.6 Myr after CAIs formed would thus incorporate enough live $^{26}$Al ($^{26}$Al/$^{27}$Al > 4 x 10$^{-6}$) to completely differentiate, like Vesta. Asteroids that grew this large from 2.6 Myr - 4.5 Myr after the formation of CAIs at the beginning of the Solar System would incorporate less (4 x 10$^{-6}$ > $^{26}$Al/$^{27}$Al > 5 x 10$^{-7}$) live $^{26}$Al, enough to melt water ice but not rock. Such asteroids would be abundant in products of aqueous alteration, in particular phyllosilicates. If peak temperatures in the asteroid exceeded 400 °C, these phyllosilicates would decompose and devolatilize. Such asteroids would resemble S-type asteroids, the presumed parent bodies of ordinary chondrites. If peak temperatures did not exceed 400º C, asteroids in which water melted would retain phyllosilicates and resemble C-type asteroids, the presumed parent bodies of carbonaceous chondrites. Support for heating as the cause of devolatilization comes from the observed dichotomy inherent in the van Schmuss-Wood classification scheme of chondrites; that they either retained abundant hydrated phases (petrologic types 1 and 2) *or* were heated above 400º C (petrologic types 3-6) [van Schmus & Wood 1967, Weisberg *et al.* 2006]. Petrologic evidence from CV chondrites strongly indicates they devolatilized by heating (Krot *et al.* 1995; Kojima & Tomeoka 1996). In asteroids that incorporated even less live $^{26}$Al ($^{26}$Al/$^{27}$Al < 5 x 10$^{-7}$), ice would never melt, and phyllosilicates would not be produced. These asteroids presumably would probably resemble the D- and P-type asteroids of the outer asteroid belt.

Radiogenic heating by $^{26}$Al is a major control on planetesimal volatile inventory (Gilmour & Middleton, 2009).

GM93 argue that the heliocentric zoning of the asteroid belt is a result of varying amounts of incorporated live $^{26}$Al, due to increasing accretion times with increasing distance from the Sun. Asteroids inside 2.7 AU grew quickly enough (< 2.6 Myr) to retain sufficient $^{26}$Al to climb above 400° C, and now resemble the S-type asteroids. From 2.7 to 3.4 AU, less $^{26}$Al was retained, the asteroids did not devolatilize, and C-type asteroids were produced. Beyond 3.4 AU, so little live $^{26}$Al was incorporated by the time these planetesimals grew large, that ice in these bodies never melted, producing D- and P-type asteroids. These zones are quite narrow, and are not dependent on the amount of radiation received from the Sun (the internal temperatures of asteroids due to $^{26}$Al decay, > 700 K, are little affected by the Sun's luminosity).

Phyllosilicates apparently are produced on asteroids with just enough $^{26}$Al to melt ice, but not so much $^{26}$Al that the asteroids devolatilize: $5 \times 10^{-7} < {}^{26}Al/{}^{27}Al < 4 \times 10^{-6}$. In our Solar System, this required the $^{26}$Al in CAIs to decay for 3-5 Myr from its initial value of $^{26}Al/^{27}Al = 5 \times 10^{-5}$. It is highly likely that this initial abundance of $^{26}$Al was fixed by the amount of material the Solar System incorporated from a nearby supernova (Lee *et al.* 1976, Hester *et al.* 2004, Jacobsen 2005, Wadhwa *et al.* 2007), either from ejecta that contaminated its molecular cloud core (Cameron & Truran 1977, Boss & Vanhala 2000), or from ejecta injected into its already-formed protoplanetary disk (Ouellette *et al.* 2005, 2007; see also Looney *et al.* 2006). Either way, injection of supernova material into a forming planetary system is a highly stochastic process, and other planetary systems are likely to have very different initial $^{26}$Al abundances. Specifically, protoplanetary disks in regions that lack a massive star (e.g., the Taurus-Auriga region) will incorporate no $^{26}$Al from a nearby supernova, and almost certainly have $^{26}$Al/$^{27}$Al ratios orders of magnitude lower than in our Solar System.

In a system with $^{26}$Al/$^{27}$Al < 5 x 10$^{-7}$, ice would not melt on any planetesimals. Any planetesimals that formed in a part of a protoplanetary disk where ice condensed would contain abundant water ice, resembling D- and P-type asteroids. (The disk temperatures in a typical disk will not exceed the sublimation temperature of ice, 180 K, outside of about 0.7 AU: Chiang & Goldreich 1997). In such systems, no phyllosilicates would be produced on any planetesimals. On the other hand, terrestrial planets that form in such systems, mostly from planetesimals inside 2.6 AU (Raymond *et al.* 2004, 2006), are likely to contain substantially more water than Earth, perhaps closer to tens of percent by weight. Planets in such $^{26}$Al-poor systems would be "waterworlds," whose internal structures were explored by Léger *et al.* (2004). $^{26}$Al-rich systems, like our Solar System, would possess planetesimals where ice melted, producing phyllosilicates. Terrestrial planets in such systems are predicted to be "dry", like Earth. Likely regions to search for phyllosilicate emission from such systems, would be protoplanetary disks in the Orion Ic and Id subgroups, whose O and Si abundances strongly suggest contamination by supernovae in the Orion Ia and Ib subgroups (Cunha & Lambert 1992, 1994, Cunha *et al.* 1998).

The hypothesis that planetesimals in star-forming regions like Taurus will be ice-rich, producing water-rich terrestrial planets, or "waterworlds", admittedly is based on a number of assumptions. One test of this "waterworlds" hypothesis is that the dust in protoplanetary disks in regions like Taurus-Auriga will not exhibit phyllosilicates. Phyllosilicates, we predict, are possible only in systems that formed near a supernova, with abundant $^{26}$Al. ***If*** the material in $^{26}$Al-rich, phyllosilicate-bearing planetesimals is returned to the protoplanetary disk, and ***if*** phyllosilicates can be spectrally distinguished, then phyllosilicate emission is predicted to be found in these $^{26}$Al-rich systems, and in these systems only.

*Mid-Infrared (MIR) Spectra*

Strong observational evidence for disks around young stellar objects (YSOs) and T Tauri stars exists, especially in the form of excess infrared (IR) emission over what would be expected from the stellar photosphere alone (McCaughrean & O'Dell, 1996; Adams *et al.*, 1988; Chiang & Goldreich, 1997; de Pater & Lissauer, 2001). In fact, IR excess emission is observed in 25-50% of pre-main-sequence stars of 1 $M_\odot$ (de Pater and Lissauer, 2002). The excess IR emission is the result of thermal emission due to reprocessed starlight from circumstellar dust grains (Adams *et al.*, 1988; Hartmann, 1998). During the T

Tauri stage, the system consists of a passive, reprocessing disk, in which the excess infrared emission originates from dust grains in the outer layers of the disk that are heated by starlight (Kenyon & Hartmann, 1987).

The spectra of YSOs with IR excesses contain information about the composition of the dust that gives rise to that emission. While observations at millimeter wavelengths probe closer to the midplane of the disk, mid-infrared observations probe the surface layers of the disk. The spectra of YSOs routinely exhibit silicate emission bands at $\lambda \sim 10$ μm and $\lambda \sim 20$ μm. This implies that the grains are emitting in the Rayleigh limit, $a \leq \lambda/2\pi$, where $a$ is the radius of the grain, as the observed silicate feature should diminish if the particle size was more than a few microns in size (Pollack, 1994; Nakamura, 1998). The silicate band positions and profiles are highly diagnostic of stoichiometry (Dorschner *et al.*, 1995; Fabian *et al.*, 2001; Krishna Swamy, 2005). Phyllosilicates exhibit the 10 μm and 20 μm features characteristic of silicates, with distinctive substructure particular to each specific mineral (Figure 1). All phyllosilicates also show a distinct absorption feature at 6 μm and other wavelengths due to structural $H_2O$. The unique and distinctive mid-infrared spectral features of silicates, in general, and phyllosilicates, in particular, make the study of the mineralogy of protoplanetary disks possible.

*The Disk Environment*

Forming planetary systems are generally observed in one of two states: the protoplanetary disk stage, when gas and dust are both present and presumed close in properties to interstellar material; and the debris disk stage, after gas has been removed and only dust from collisions between planetesimals is present. Examples of protoplanetary disks are abundant (Adams, Lada & Shu 1987) and include the archetype T Tauri. The fraction of disks in a cluster observed to possess protoplanetary disks tends to decrease with age of the cluster, dropping below 50% at 3-6 Myr (Haisch *et al.* 2001). Examples of debris disks include β Pic (Smith & Terrile 1984) and AU Mic (Kalas *et al.* 2004); both members of the 12 Myr old β Pic moving group (Zuckerman *et al.* 2001). Objects apparently in transition between the two also exist, such as TW Hya (Calvet *et al.* 2002). Since the formation of terrestrial planets takes several tens of Myr (Wadhwa & Russell 2000), ideally debris disks would be better to observe, as they would contain only dust from planetesimals, at the time terrestrial planets are forming. Unfortunately, debris disks are too faint to search for the infrared signatures of phyllosilicates. The average column density of debris disks range from ~ $10^{-4}$ to ~ $10^{-7}$ g cm$^{-2}$, as compared to the column density in the superheated dust layer of a protoplanetary disk, which is on the order of ~ $10^{-2}$ g cm$^{-2}$ (Chiang & Goldreich 1997). The dust emission features (which can arise only in the optically thin portions of disks) are therefore several orders of magnitude weaker in debris disks than in protoplanetary disks. The fluxes from the nearest debris disk, β Pictoris, are sufficiently large to detect the dominant emission features, such as crystalline and amorphous silicates (Okamoto *et al.* 2004); but as we show below, the fluxes in debris disks are too low to detect features that are a few percent on the continuum, such as the phyllosilicates we consider here. We therefore do not consider debris disks further and turn our attention to protoplanetary disks.

If protoplanetary disks contained predominantly interstellar dust, then their spectra would not yield any new information about the processes in the disk or the composition of planetesimals in them. Protoplanetary disks do not, however, contain pure interstellar dust. The existence of crystalline silicates argues strongly for thermal processing of dust within them (e.g., Wooden *et al.* 2007), as the interstellar medium contains only (> 99.8%) amorphous silicates (Kemper *et al.* 2004). Dust samples from of comet 81P/Wild 2, returned as a part of the STARDUST mission, have been shown to have a solar isotopic composition (Brownlee *et al.* 2006; Zolensky *et al.* 2008, Stephan 2008), indicating processing in the early Solar System. Observations of Orion proplyds also indicate growth of the minimum grain size in the first few x $10^5$ years (Throop *et al.* 2001). This is *not* interpreted to mean that grains took > $10^5$ years to collide. Quite the contrary, micron-sized grains coagulate and fragment on timescales < 10 years typically, rapidly establishing a quasi steady-state grain size distribution; it is the mean size of the distribution that evolves over long time periods (Throop *et al.* 2001).

Numerical models of grain growth, from sub-micron to planetesimals tens of kilometers in diameter, also predict it to be rapid: $< 10^4$ yr at 3 AU (Woolum & Cassen 1999; Weidenschilling 2000; Weidenschilling & Cuzzi 2006).  In fact, the basaltic howardite, eucrite and diogenite achondrites (spectrally associated with the 200-km diameter asteroid, 4 Vesta: Drake 1979), formed on a fully differentiated body.  Al-Mg isotopic studies show their parent body differentiated *and crystallized* within at most 5 Myr after the formation of the first solids in the Solar System, CAIs (Srinivasan *et al.* 1999).  Bodies tens of kilometers in diameter that formed within the first 4 Myr of our Solar System contained sufficient $^{26}$Al to melt water ice within them (Grimm & McSween 1993), potentially leading to phyllosilicate production on these bodies by processes such as serpentinization (Cohen & Coker 2000).  If similarly sized bodies are also forming in protoplanetary disks, and if these bodies are also collisionally ground to dust during this stage, much of the dust seen in protoplanetary disks could be a mix of interstellar dust that has coagulated and been processed in the nebula, plus dust shed by planetesimals, akin to zodiacal dust.

Numerical simulations of the coagulation of dust and accretion of larger bodies, as well as the collisional disruption of bodies, strongly suggest that at a few AU planetesimals are built up and torn down on timecales < 1 Myr (Weidenschilling 2000).  At any one time, about half of the mass of solids at a few AU will reside in (roughly micron-sized) dust grains, the other half residing in larger planetesimals (Weidenschilling 1997, 2000; Dullemond & Dominik 2005; Brauer *et al.* 2008; Johansen *et al.* 2008). The dust is sub-micron in size because that is the size distribution arising from a balance between coagulation and fragmentation. Intriguingly, it matches the sizes of matrix grains in chondrites (Brearley 1996). This dust is *not* interstellar dust that simply has not yet had time to accrete into planetesimals. Rather, accretion models would say it is primarily dust eroded from the larger planetesimals in the disk.

A simple analysis of asteroidal erosion also yields the same conclusion, that dust is primarily derived from larger bodies. In the current Solar System, there is approximately $10^{20}$ g of zodiacal dust shed from asteroids (Nesvorny *et al.* 2006) or the sublimation of comets (Lisse 2002). Assuming a typical grain radius of 10 micron, these dust grains have a lifetime against Poynting-Robertson drag $\cong$ 0.5 Myr (Wyatt & Whipple 1950), implying a production rate of up to 2 x $10^{14}$ g yr$^{-1}$, due to asteroid-asteroid collisions. It is understood that the present-day asteroid belt is *highly* depleted relative to the primordial asteroid belt, by a factor of about $10^3$ (Weidenschilling 1977, Bottke *et al.* 2005). If the primordial asteroid belt contained 1000 times the number of asteroids as it does today, and assuming the same velocity dispersions, the production rate of dust (proportional to the number density squared) would have been 2 x $10^{20}$ g yr$^{-1}$. Pumping of asteroid eccentricities by a primordial Jupiter (Weidenschilling *et al.* 2001) could have increased the collision rate, and the production rate, even more. Collisions between planetesimals in protoplanetary disks should generate as much dust as is present in the Solar System zodiacal dust, *every year*.

Unlike dust in a debris disk, dust in a protoplanetary disk is constrained to follow the gas and will not be lost to Poynting-Robertson drag. After $10^6$ yr of collisions between asteroids, one might reasonably expect roughly $2 \times 10^{26}$ g of dust to be shed from a primordial asteroid belt. This is to be compared to our assumed mass of the primordial asteroid belt itself, 1000 times the current mass of $3 \times 10^{24}$ g, or about $3 \times 10^{27}$ g (Bottke *et al.* 2005). That is, something on the order of 7% of the nebular dust that grows to planetesimal size could be returned to the nebula gas as zodiacal dust on Myr timescales. This mass in asteroids roughly equaled the mass in dust at the same time, so for the innermost few AU of a protoplanetary disk at least a few $\times 10^5$ yr old, then, it is reasonable to conclude that a significant fraction, $\approx 7\%$, of the dust is eroded from planetesimals already formed in the disk.

If planetesimals in a protoplanetary disk have had time to grow and retain radioactive heat sufficient to melt water ice, the dust eroded from them could contain products of aqueous alteration. These products include carbonates and especially phyllosilicates (Mottl *et al.* 2007). We focus on phyllosilicates because they are the most common products of aqueous alteration found in carbonaceous chondrites and interplanetary dust particles (IDPs). The zodiacal dust in our own Solar System, derived from comets and asteroidal collisions, has a spectrum consistent with the inclusion of 20% abundance of the phyllosilicate montmorillonite (Reach *et al.* 2003). As discussed below, phyllosilicates can comprise a large fraction of the mass in carbonaceous chondrites (40-90%). Assuming a mass fraction on the order of 50%, we expect that 3% of the dust in the nebula could be composed of phyllosilicates, especially in the regions (2-4 AU) that correspond to the location of the Solar System's asteroid belt. These phyllosilicates should radiate most strongly at ~ 20 microns.

*Phyllosilicates*

Phyllosilicates are the major mineral product of aqueous alteration of silicate rock, and therefore are excellent tracers of liquid water. Aqueous alteration on the parent body is believed to account for the majority of the phyllosilicates found in meteorites (Krot *et al.* 2006), although fine-grained rims around chondrules may have been produced in the gas phase (Ciesla *et al.* 2003). The water contained in carbonaceous chondrites is mainly in the form of these hydrous minerals. In the subclass of CI carbonaceous chondrites, anywhere from 40% to over 90% of the volume is made up of fine-grained phyllosilicates and associated phases (Tomeoka & Buseck 1990; Buseck & Hua 1993; Rubin 1997). Phyllosilicates are also often found in the fine-grained rims around chondrules (Ciesla *et al.* 2003 and references therein).

Virtually all carbonaceous chondrites contain phyllosilicates, with the different amounts indicating varying degrees of aqueous alteration. The subclasses which have experienced the most alteration are the CI, CM, and CR chondrites (Buseck & Hua 1993; Rubin 1997), which are significantly more altered than the CO and CV chondrites. The most common phyllosilicates found in meteorites (Table 1) are overwhelmingly saponite ($[Ca/2,Na]_{0.33}[Mg,Fe^{2+}]_3[Si,Al]_4O_{10}[OH]_2 \cdot 4H_2O$) and serpentine ($[Mg,Fe]_3Si_2O_5[OH]_4$), with montmorillonite ($[Na,Ca]_{0.33}[Al,Mg]_2Si_4O_{10}[OH]_2 \cdot n[H_2O]$) identified in the matrix of some CI chondrites and in the fine-grained rims of chondrules (Buseck & Hua 1993). Cronstedtite ($Fe_2^{2+}Fe^{3+}[Si,Fe^{3+}]O_5[OH]_4$), a Fe-rich phyllosilicate (Lauretta 2000; Buseck & Hua 1993) rarely found terrestrially, makes up the bulk of the matrix of CM chondrites, along with intergrowths of serpentine (Buseck & Hua 1993).

*Outline*

This paper is organized as follows. We first review the minerals expected in the "zodiacal dust" in protoplanetary disks, i.e., those found in meteorites. We then discuss use of Mie theory to convert the complex index of refraction measured for these minerals into opacities. We then use a simple 2-layer radiative transfer model to predict the spectral energy distributions (SEDs) from protoplanetary disks of varying mineralogies. We present our model SEDs with and without the inclusion of 3% phyllosilicates and show that the differences in the spectra are observable with many infrared instruments. We conclude that phyllosilicates are detectable by their infrared emission in protoplanetary disks.

Finally, we discuss use of phyllosilicate emission to test the "waterworlds" hypothesis that suggests that the water content of planetesimals and terrestrial exoplanets is linked to the abundance of short-lived radionuclides such as $^{26}$Al (Desch & Leshin 2004).

**METHODS**

*Minerals Expected*

The dust in protoplanetary disks is expected to consist primarily of compounds of the rock-forming elements (Si, O, Fe, Mg, Al, Ca, Na, S, Ni) along with carbonaceous compounds. Based on cosmic element abundances, the most abundant species should be compounds of Si, O, Mg, and Fe, plus carbon dust (Gail, 2003). Silicates are expected to be an amorphous non-equilibrium mixture in the outer portion of the disk, progressing through a crystalline non-equilibrium mixture in the middle portion of the disk, to a crystalline equilibrium mixture in the inner portion of the disk (Gail, 2003). Gail (2003, 2004) used the grain model of Pollack *et al.*, (1994) to arrive at an outer disk composition of amorphous olivine, pyroxene, and quartz, solid iron and troilite, and kerogen. Kerogen was chosen as representative of the carbon-rich component of Pollack *et al.*, (1994), as it is the carbonaceous material found in the matrix of carbonaceous chondrites (Gail, 2003). The relative abundances of these species are shown in Table 2. Gail (1998, 2003, 2004) estimates that the dust grains of the inner regions of the disk consist of pure, crystalline forsterite ($Mg_2SiO_4$) and enstatite ($MgSiO_3$), with the innermost regions iron and carbon-free, yet containing an aluminum component.

Based on the minerals expected in disks, the minerals considered in this study include both amorphous and crystalline olivines and pyroxenes (forsterite and enstatite), troilite (FeS), quartz ($SiO_2$), hibonite ($CaAl_{12}O_{19}$), and the phyllosilicates most commonly found in meteorites: saponite, serpentine, and montmorillonite. This list is hardly exhaustive and does not include amorphous carbon or other sources of continuum opacity. Inclusion of such minerals would be important if one were fitting an observed SED. Our purpose here is to compare differences in computed spectra with and without phyllosilicates, so neglect of these other, relatively featureless minerals should not alter our conclusions.

*Optical Properties*

In order to properly model and interpret the spectral energy distributions (SEDs) of protoplanetary disks, it is necessary to understand how the small dust particles in the surface layers of the disk interact with the radiation received from the central star. We consider the case in which the particles are small compared to the wavelength of radiation (within the Rayleigh limit). This is the appropriate limit needed for the production of observable mid-infrared silicate features (Pollack *et al.*, 1994; Nakamura 1998).

Dust particles in an accretionary disk of a YSO will result in the extinction of radiation emitted by the protostar by absorption and scattering. Extinction strongly depends on the size, shape, and chemical composition of the particles (Bohren & Huffman, 1983; Min *et al.*, 2003). The scattering and absorption properties of homogeneous spherical particles can be determined extremely accurately utilizing Mie theory, but dust grains are decidedly no spherical. Mie theory can also be employed to calculate opacities of a population of dust grains with particular size or shape distributions (such as spheroids or ellipsoids), albeit with significant caveats (Bohren & Huffman, 1983; Min *et al.*, 2003).. Some of these distributions provide decent approximations to the opacities of real powders or other distributions of particles.

Min *et al.* (2003) performed an extensive study of the shape effects in scattering and absorption by small particles. They found that all shape distributions other than homogeneous spheres matched the position of maxima from astronomical objects (circumstellar dust). A distribution of hollow spheres (DHS) provided the overall best fit to the observations in the Min *et al.* (2003) study, leading to our decision to employ this shape distribution in the present study.

*Protoplanetary Disk Model*

The disk model presented in this study is based on that of Chiang & Goldreich (1997). They present a model of a passive, flared disk in hydrostatic and radiative equilibrium (Figure 2). We too, only consider those cases in radiative equilibrium, where the radiation emitted from a particle of dust is equal to the radiation absorbed. In this study, however, we have calculated dust opacities using a Distribution of Hollow Spheres and a population of grains of different compositions, as opposed to the simple opacity approximation used by Chiang & Goldreich (1997). We also corrected an error of a factor of $2^{1/4}$ in the equation for effective temperature given by Chiang & Goldreich (1997). In addition, Kurucz (1993) models of stellar atmospheres were used, rather than simply modeling the central star as a blackbody.

The parameters of our model include a T Tauri star of effective surface temperature, $T_* = 6000$ K, mass, $M_* = 1$ M$_\odot$, and radius, $R_* = 1$ R$_\odot$, with a passive, flared, reprocessing disk (face-on). The distance to the star was chosen to be 145 parsecs (approximately the distance to the Taurus-Auriga star-forming region). The inner radius of the disk is located 2 astronomical units (AU) from the central star, and the outer radius of the disk was set at 80 AU. The model SED resulting from these parameters is shown in Figure 3.

**RESULTS**

The composition of the dust and the relative percentages of each mineral used in our model are shown in Table 3. Opacities were calculated based upon a grain size of 0.1 μm, because we expect the dust in protoplanetary disks to exist in a quasi steady-state size distribution, with the smallest particles being submicron in size (Weidenschilling 2000; Dullemond & Dominik 2005). The presence of submicron dust, even after many Myr of disk evolution, is strongly argued for: observations of the 10 μm silicate feature suggest that grains smaller than 2 μm in radius exist for several Myr in disks (van Boekel et al. 2004; Dullemond & Dominik 2005). Micron-sized dust also exists in the form of matrix grains in chondrites that formed many Myr after CAIs formed (Brearley 1996). Additionally, although composite aggregates are expected in protoplanetary disks, it is customary to model them by assuming collections of smaller particles of various composition (van Boekel *et al.* 2005; Min *et al.* 2008). The effect of phyllosilicates on the spectra was tested by replacing 3% (as derived earlier; see *The Disk Environment*) of the amorphous forsterite with an equal amount of saponite (the most common phyllosilicate found in the matrix of meteorites).

The difference in the SED resulting from the inclusion of 3% phyllosilicates is certainly evident, as shown in Figure 4. To ensure that the difference observed was due to the inclusion of phyllosilicates, rather than the decrease in the amount of amorphous forsterite, model SEDs were produced with 75% amorphous forsterite and 12% amorphous enstatite in one case and 12% amorphous forsterite and 75% amorphous enstatite in another (with all other relative percentages remaining the same as in Table 3). The difference between the SEDs with and without the inclusion of 3% phyllosilicates was still evident.

Inspection of Figure 4 shows that the overall flux is increased with the inclusion of phyllosilicates, with the major spectral features located at the same wavelength as without phyllosilicates. At first glance, it would seem that it would be difficult to distinguish between an SED with or without phyllosilicates. While there are measurable differences in the calculated models with and without phyllosilicates, broad differences in level can be difficult to detect in the spectra of astrophysical objects due to uncertainties in the exact shape of the underlying continuum and foreground extinction and screening, as well as the difficulty of doing absolute spectrophotometry. The sensitivities of instruments in the infrared are rarely known better than a few %, because of the difficulty of calibration, and the fact that they do not counting photons. In practical settings, detection is both easier and more reliable if it involves comparison of distinct features that can be isolated from the background. Upon closer examination of Figure 4, one can see higher emission at ~ 21 μm than ~ 24 μm in the SED including phyllosilicates. In the SED without phyllosilicates, however, the reverse is true; the ~ 24 μm emission is stronger than the ~ 21 μm emission. The ratio of these two features allows one to approach the question of detectability quantitatively. As the model SED shown in Figure 4 is representative of the inclusion of saponite, we first wish to determine if the same trend is seen when including different species of phyllosilicates. Model SEDs with the inclusion of 3% serpentine, montmorillonite, or cronstedtite are shown in Figure 5. For the model SED with the inclusion of saponite, the flux ratio, $F_{21}/F_{24} = 1.093$, as opposed to the model SED without the inclusion of phyllosilicates, $F_{21}/F_{24} = 0.882$. The model SEDs with the inclusion of serpentine, montmorillonite, and cronstedtite show $F_{21}/F_{24} = 1.065$, $F_{21}/F_{24} = 1.282$, and $F_{21}/F_{24} = 0.872$ respectively. It appears that the flux ratio, $F_{21}/F_{24}$, may be diagnostic of the presence of saponite, serpentine, and montmorillonite, although is not diagnostic of the presence of cronstedtite. Figure 6 shows a "color-color" plot of the model SEDs.

*Possibility for Detection*

The question remains whether the detection of the effects of phyllosilicates on the 21 μm / 24 μm ratio is possible with the instruments available (or soon to be available). In the model SED including saponite, the flux at 21 μm is ~ 4250 mJy and the flux at 24 μm is ~ 3889 mJy. This is well above the sensitivities for the instruments under consideration (see Table 4). Based upon the model SED (with the inclusion of saponite), $R = 10.4$, $F_{\lambda_1} = 4250$ mJy, and $F_{\lambda_2} = 3889$ mJy, we have determined the minimum instrument-specific integration time, $t$, necessary to achieve a 1σ detection of $\mathcal{R} = F_{\lambda_1} / F_{\lambda_2}$, (see Appendix A):

$$t = \left(\frac{\Sigma_{\mathcal{R}}}{\Sigma_0}\right)^2 \left(\frac{F_0}{\sqrt{F_{\lambda_1} F_{\lambda_2}}}\right) \left(\frac{R}{R_0}\right) t_0. \qquad (11.)$$

Table 4 shows *t* for the Spitzer Space Telescope, the Stratospheric Observatory for Infrared Astronomy (SOFIA), Michelle (Gemini North), the NASA Infrared Telescope (IRTF), and the James Webb Space Telescope (JWST). The question remains as to whether the calculated minimum integration time is reasonable.

The minimum integration times for the Spitzer Space Telescope, the Stratospheric Observatory for Infrared Astronomy (SOFIA), and the James Webb Space Telescope (JWST) indicate it is possible to achieve a 1σ detection. For Michelle (Gemini North), the minimum integration time of ~ $10^3$ s may be reasonable; however, background noise and dark current will become significant over that integration time. A closer examination of the noise contributions would be necessary to determine definitively if it is possible to achieve a 1σ detection. A more careful analysis of the noise is also necessary for the NASA Infrared Telescope, as background and dark current may become problematic with an integration time of 10 minutes. The demonstration that phyllosilicates can be detected with the Spitzer Space Telescope is of particular value, considering the availability of large amounts of archived data on protoplanetary disks as a part of the Spitzer Legacy Science Program.

**DISCUSSION**

The implications of detection of phyllosilicates within protoplanetary disks are immense. Since phyllosilicates are formed by the interaction of liquid water with rock on planetesimals, their presence would strongly point to the existence of liquid water on rocky bodies in such protoplanetary disks. Detection of phyllosilicates would represent the first, albeit indirect, discovery of liquid water in another planetary system. The presence of liquid water has obvious astrobiological implications, both for the commonality of systems like our own, as well as the development of life elsewhere. Additionally, their abundance in the dust of the protoplanetary disk would verify numerical models (e.g., Weidenschilling 2000) that suggest strong recycling of material between the dust of the disk and the planetesimals in the disk.

If phyllosilicates can indeed be reliably detected in protoplanetary disks, sufficient statistics may be accumulated to probe the abundances of the short-lived radionuclide $^{26}$Al in other planetary systems. The "waterworlds" hypothesis put forth by Desch & Leshin (2004) first proposed that water content in terrestrial planets forming in a solar system is linked to the $^{26}$Al abundance in that system, to the star-forming environment in which that system formed, and to observational signatures such as phyllosilicate emission. A prediction of the waterworlds hypothesis is that disks in the Taurus-Auriga molecular cloud, which have not been exposed to supernova ejecta and which therefore should contain low $^{26}$Al ($^{26}$Al/$^{27}$Al $\ll 5 \times 10^{-7}$), will not exhibit phyllosilicate emission, but would also be predicted to have water-rich terrestrial planets. On the other hand, disks in systems that were contaminated by supernova material, such as those in the Orion Ic and Id associations (Cunha & Lambert 1992, 1994; Cunha *et al.* 1998) are likely (but will not necessarily) contain abundant $^{26}$Al and produce planetesimals containing phyllosilicates disks in regions such as Taurus-Auriga.

*Future Work*

In addition to phyllosilicates, carbonates are also important products of aqueous alteration (Mottl *et al.* 2007). As well as their presence in meteorites, Lisse *et al.* (2006) also tentatively identified the carbonates magnesite ($MgCO_3$) and siderite ($FeCO_3$) as components of the spectrum of the Deep Impact ejecta. Carbonates have strong emission features between 6-7 μm, outside of the 10 μm silicate emission band (Lane & Christensen 1997). This lack of interference from silicates may enable the easy detection of carbonates. Future modeling will investigate this possibility by producing model SEDs with and without the inclusion of carbonates.

In the future, it would be worthwhile investigating the spectral features and detectability of phyllosilicates that are less abundant. Lisse et al. (2006) tentatively identified nontronite in the spectra of comet 9P/Tempel 1. It would be appropriate to consider this and other phyllosilicates. For comparison with upcoming *Herschel* observations, it would also be highly appropriate to measure opacities at long wavelengths, out to 100 microns.

More sophisticated methods exist to model the absorption of radiation by particles, which have been shown to account for shape effects better than standard Mie theory. These include Coupled Dipole Approximation (CDA) and Discrete

Dipole Approximation (DDA). These methods need to be utilized to produce model SEDs, with and without phyllosilicates, to compare to our results reported here.

**CONCLUSION**

In summary, phyllosilicates in the dust of protoplanetary disks imparts an observable signature, especially in the 21-24 micron region. Fitting of model SEDs to disks observed by infrared observatories, including the Spitzer Space Telescope, could resolve whether phyllosilicates are present at the few percent level. We expect phyllosilicates to be present in disks at levels such as these if formation and collisional destruction is rapid compared to disk evolution timescales, as predicted by numerical models, and if those planetesimals contain liquid water. The detection of phyllosilicates by the means outlined here would represent the first, albeit indirection, detection of liquid water on rocky bodies in planetary systems.

**ACKNOWLEDGEMENTS**

Optical constants for amorphous forsterite and enstatite, and crystalline enstatite were taken from Jäger *et al.* (2003). Optical constants for crystalline olivine were taken from Fabian *et al.* (2001). Optical constants for troilite, quartz, and hibonite, were taken from Begemann *et al.* (1994), Henning & Mutschke (1997), and Mutschke *et al.* (2002), respectively. Optical constants for the phyllosilicates (saponite, serpentine, montmorillonite, and cronstedtite) were obtained by Timothy Glotch (Glotch 2006, personal communication; Glotch *et al.*, 2007).

This work was supported by NASA Origins of Solar Systems grant NNG06GI65G. We heartily thank Jeff Hester and Rolf Jansen for assistance on the observational aspects of the problem.

**AUTHOR DISCLOSURE STATEMENT**

No competing financial interests exist.

| Saponite   | Serpentine | Montmorillonite | Cronstedtite |
|------------|------------|-----------------|--------------|
| CI         | CM         | CI              | CM           |
| CV         | CO         |                 |              |
| CR         | CR         |                 |              |
| ord. chond.|            |                 |              |
| IDPs       |            |                 |              |

**Table 1.** Phyllosilicates found in meteorites.

| Species | Mg | Fe | Si | S | C |
|---|---|---|---|---|---|
| olivine-type | 0.83 | 0.42 | 0.63 | | |
| pyroxene-type | 0.17 | 0.09 | 0.27 | | |
| quartz-type | | | 0.10 | | |
| Iron | | 0.10 | | | |
| Troilite | | 0.39 | | 0.75 | |
| Kerogen | | | | | 0.55 |

**Table 2.** Dust composition in the outer disk. From the model of Gail (2003, 2004)

|                      | Mineral Percentages     |                      |
|----------------------|-------------------------|----------------------|
|                      | **Without Phyllosilicates** | **With Phyllosilicates** |
| amorphous forsterite | 58                      | 55                   |
| amorphous enstatite  | 32                      | 32                   |
| crystalline olivine  | 3                       | 3                    |
| crystalline enstatite| 2                       | 2                    |
| FeS (troilite)       | 2                       | 2                    |
| Quartz               | 2                       | 2                    |
| Hibonite             | 1                       | 1                    |
| Saponite             | 0                       | 3                    |

**Table 3.** Relative percentages of the minerals used in modeled SEDs

| Telescope | Instrument | R[1] | Sensitivity[2] | t[3] |
|---|---|---|---|---|
| Spitzer | IRS | 600 | 0.4 mJy | 21.8 s |
| SOFIA | EXES | 3000 | 2.7 Jy | 59.8 s |
| Gemini North | Michelle | 110 | 14 mJy | 920.9 s |
| IRTF | MIRSI | 100 | 100 mJy | 383.8 s |
| JWST | MIRI | 3000 | $5 \times 10^{-20}$ Wm$^{-2}$ | $8.3 \times 10^{-5}$ s |

**Table 4.** [1]Spectral resolution, R = $\lambda/\Delta\lambda$, is given at the relevant wavelengths (21 and 24 μm). [2]Sensitivities listed for IRS, EXES, Michelle, MIRSI, and MIRI are, 1σ for an integration time of 512s, 4σ for an integration time of 900s, 5σ for an integration time of one hour, 1σ for an integration time of 60s, and 10σ for an integration time of 10,000s. [3]Minimum integration times necessary to achieve a 1σ detection of $\mathscr{R}$.

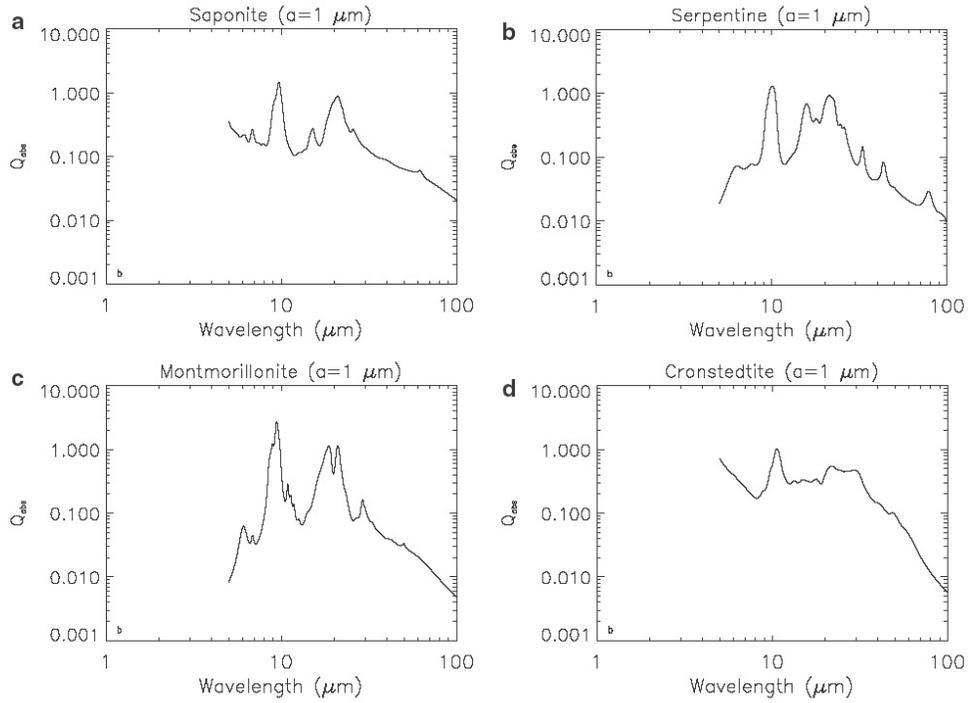

**FIG. 1.** The absorption efficiency factor, $Q_{abs}$, for (**a**) saponite, (**b**) serpentine, (**c**) montmorillonite, and (**d**) cronstedtite calculated from $n$ and $k$ determined by Glotch *et al.* (2007).

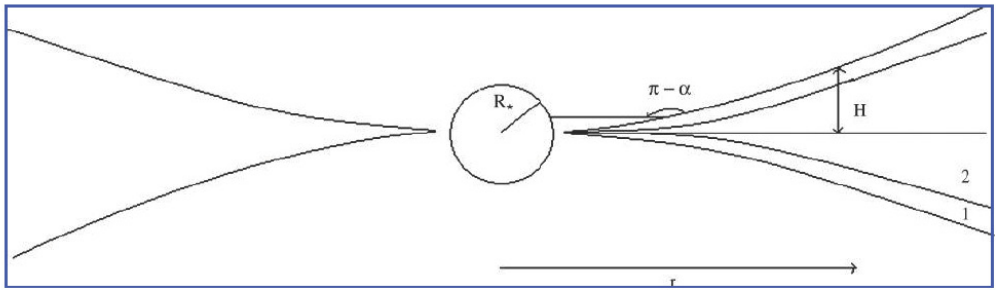

**FIG. 2.** Passive, reprocessing disk, where α is the grazing angle, *H* is the height of the visible photosphere, *r* is the distance from the star to the disk, area 1 is the superheated dust layer, and area 2 is the disk interior.

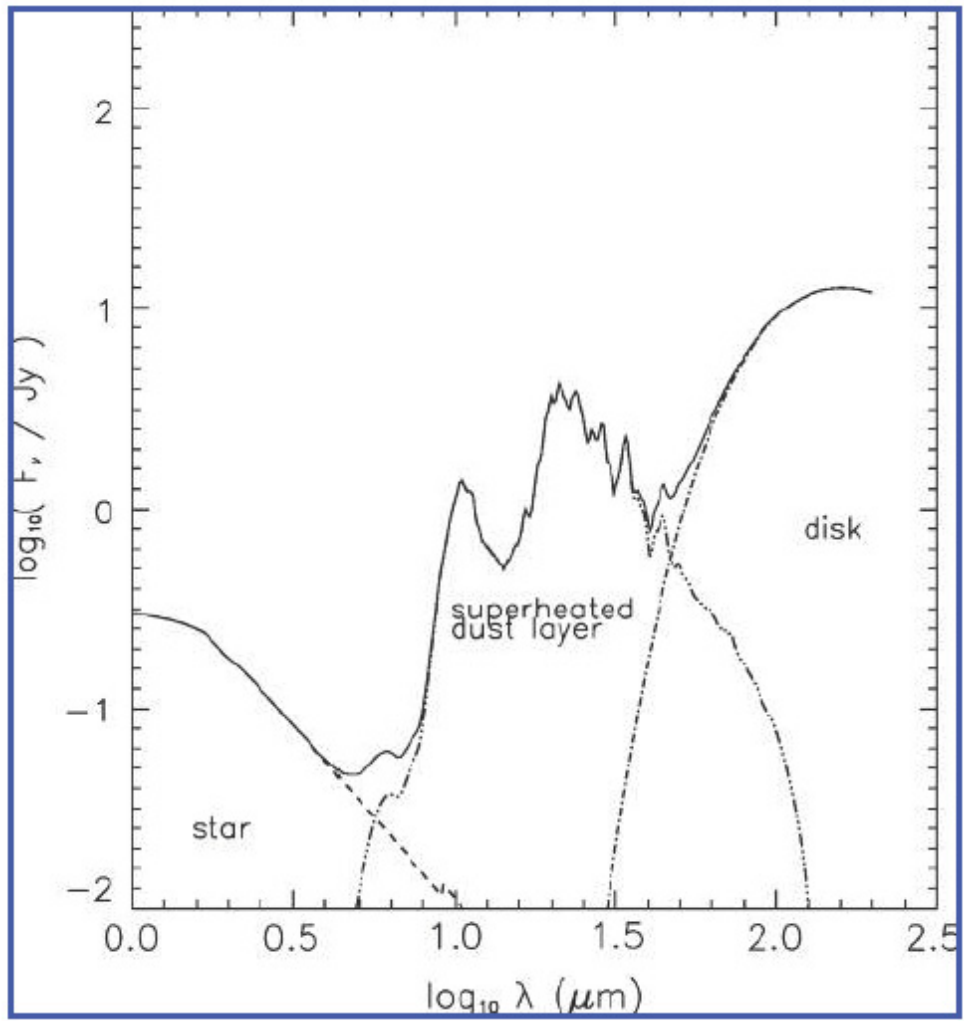

**FIG. 3.** Model SED of a passive, flared, radiative equilibrium disk, showing contributions from the star, the disk, and the superheated dust layer.

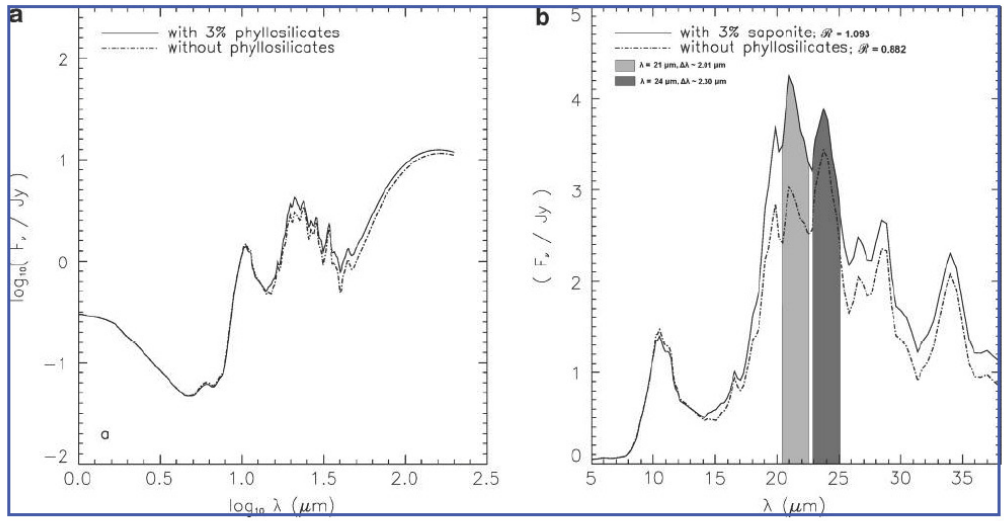

**FIG. 4.** (a) Model SED with and without the inclusion of 3% phyllosilicates. (b) Close-up view of the area of the SED where the contribution from the superheated dust layer dominates.

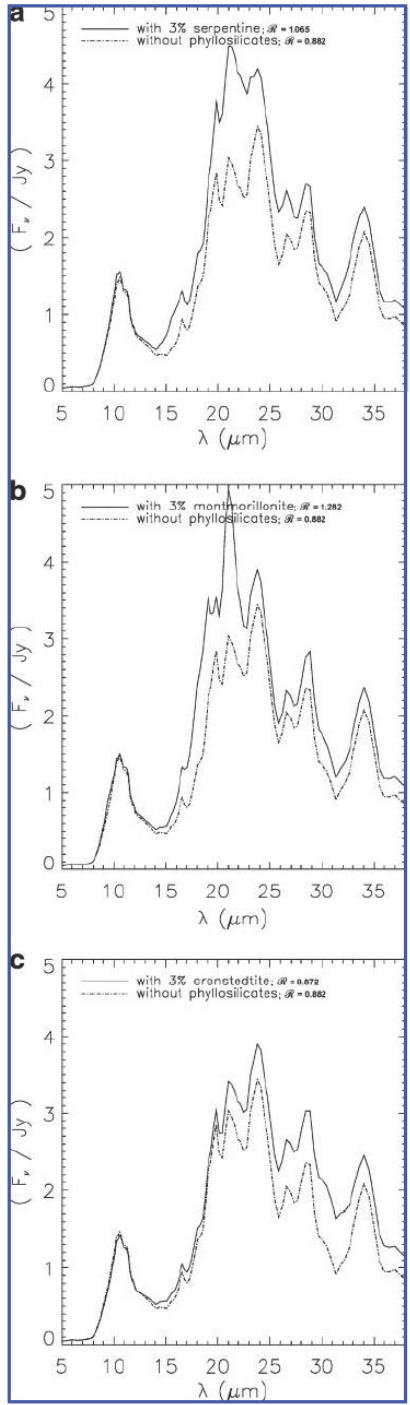

**FIG. 5.** Difference between the modeled spectrum with and without 3% (**a**) serpentine, (**b**) montmorillonite, or (**c**) cronstedtite.

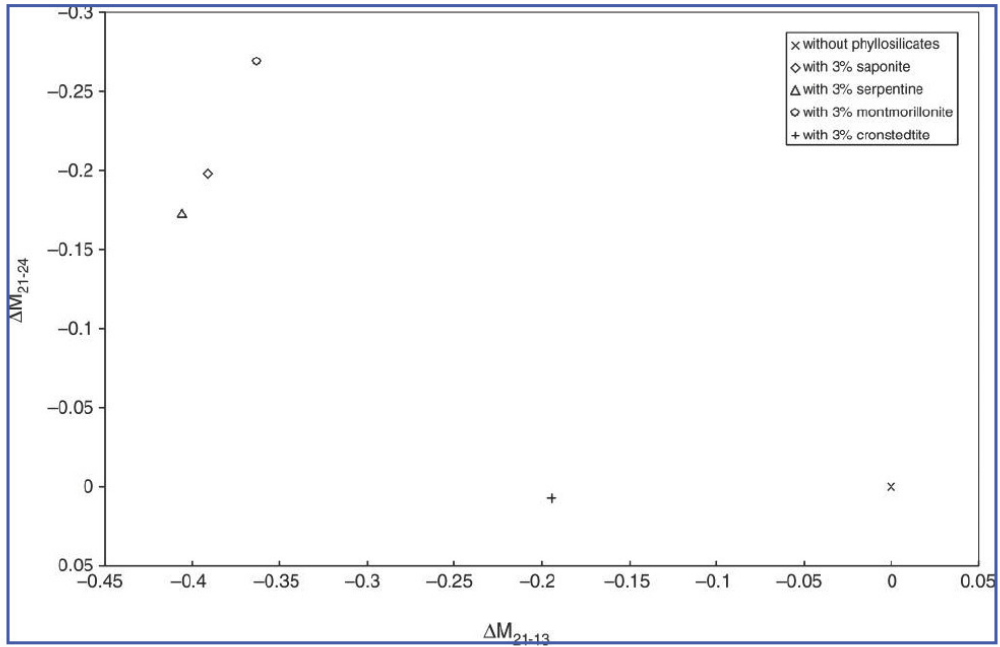

**FIG. 6.** "Color-color" diagram of the model fluxes, both with and without the addition of phyllosilicates. $\Delta M_{21-24} = -2.5 \log_{10}(F_{21}/F_{24})$ and $\Delta M_{21-13} = -2.5 \log_{10}(F_{21}/F_{13})$. The addition of phyllosilicates increases the $F_{21}/F_{24}$ ratio (except in the case of cronstedtite) in approximately the same proportion that the $F_{21}/F_{13}$ ratio increases. One can infer from this plot that phyllosilicates will show an increased flux at 21 $\mu$m.

**Appendix A**

The noise present in the measured flux, $F_\lambda$, without the inclusion of phyllosilicates is given by

$$\sigma^2_{F_\lambda} = Nt(I + S + D) + N\sigma^2_R, \qquad (A1.)$$

where $N$ is the number of pixels on which the signal is recorded, $I$ is the intensity of the signal (in electrons/pixel) at a given wavelength, $S$ is the background noise from the sky (in electrons/pixel/second; the conversion from the brightness, in ergs cm$^{-2}$ s$^{-1}$ μm$^{-1}$, involves the aperture and sensitivity of the instrument used to measure the signal), $D$ is the dark current, $t$ is the integration time, and $\sigma_R$ is the read noise. The noise in the flux measured after the inclusion of phyllosilicates is

$$\sigma^2_{F_\lambda + \Delta F_\lambda} = Nt[(I + I\delta) + S + D] + N\sigma^2_R, \qquad (A2a.)$$

where $\delta = \Delta F_\lambda / F_\lambda$. Recalling that noise adds in quadrature, this gives the total noise in the difference as

$$\sigma^2_{\Delta F_\lambda} = \sigma^2_{F_\lambda} + \sigma^2_{F_\lambda + \Delta F_\lambda} = 2Nt(I + S + D) + NI\delta t + 2N\sigma^2_R. \qquad (A2b.)$$

Absorbing the dark current into the background noise, the wavelength-dependent signal-to-noise (S/N) in the difference, $\Sigma_\Delta$, is then given by

$$\Sigma_\Delta = \frac{N(I\delta)t}{\sqrt{N}[2(I+S)t + 2\sigma^2_R + \delta It]^{1/2}}. \qquad (A3.)$$

Detectability thresholds for instruments are given in terms of the time, $t_0$ needed to detect a signal of strength $I_0$ with signal-to-noise, $\Sigma_0$.

$$\Sigma_0 = \frac{N_0 I_0 t_0}{\sqrt{N_0}\left[I_0 t_0 + S_0 t_0 + \sigma_R^2\right]^{1/2}}. \qquad (A4.)$$

It is now possible to calculate what integration time is necessary to achieve the threshold S/N, $\Sigma_0$, of a number of instruments.

The spectral resolution, $R = \lambda/\Delta\lambda$, required for detection of the features at ~ 21 μm and ~ 24 μm is determined by measuring the full width at half maximum (FWHM) of the feature at the given wavelength. $R_0$ is the spectral resolution reported on each instrument. The number of pixels over which the signal is recorded (the dispersion) must also be taken into account when determining spectral resolution. This results in the relationship $N/N_0 = R_0/R$, where N = $\Delta\lambda_R$/pixels, $N_0 = \Delta\lambda_{R_0}$/pixels. Using this relationship gives $N = N_0(R_0/R)$. As $I/I_0 = F_\lambda/F_0$, $I = (F_\lambda/F_0)I_0$, where $F_0$ is the flux measured at the given wavelength by the instrument (the sensitivity). Recall that $\delta = \Delta F_\lambda/F_\lambda$, which tells us $\Sigma_\Delta \propto \Delta F_\lambda / \sqrt{2F_\lambda}$. Making the assumptions that $\sqrt{I} \gg \sigma_R^2$ (shot-noise limited), $\delta \ll 1$ (the difference in emission due to phyllosilicates is small compared to the total emission without phyllosilicates), the source is bright compared to the background ($S \ll I_0$), and that the dark current is negligible ($D \ll I_0$) gives

$$\Sigma_\Delta = \frac{\Delta F_\lambda}{2 F_\lambda F_0}\left(\frac{R_0}{R}\right)^{1/2}\left(\frac{t}{t_0}\right)^{1/2}\Sigma_0. \qquad (A5.)$$

For a 1σ detection of the ratio, $\mathcal{R} = F_{\lambda_1} / F_{\lambda_2}$

$$\left(\frac{R_0}{R}\right)^{1/2}\left(\frac{t}{t_0}\right)^{1/2}\left(\frac{\sqrt{F_{\lambda_1}F_{\lambda_2}}}{F_0}\right)^{1/2}\Sigma_0 \geq \Sigma_{\mathcal{R}} \qquad (A6.)$$

where $\Sigma_{\mathcal{R}}$ is the signal-to-noise in the ratio. Solving for $t$ gives the minimum instrument-specific integration time necessary to achieve a 1σ detection of $\mathcal{R}$:

$$t = \left(\frac{\Sigma_{\mathcal{R}}}{\Sigma_0}\right)^2 \left(\frac{F_0}{\sqrt{F_{\lambda_1}F_{\lambda_2}}}\right)\left(\frac{R}{R_0}\right)t_0. \qquad (A7.)$$

Raymond, S. N., Quinn, T., and Lunine, J. I. (2004) Making other earths: dynamical simulations of terrestrial planet formation and water delivery. *Icarus*, 168, 1-17.

Raymond, S. N., Quinn, T., and Lunine, J. I. (2006) High-resolution simulations of the final assembly of Earth-like planets I. Terrestrial accretion and dynamics. *Icarus*, 183, 265-282.

Rubin, A. E. (1997) Mineralogy of meteorite groups. *Meteoritics and Planetary Science*, 32, 231-247.

Selsis, F., Kasting, J. F., Levrard, B., Paillet, J., Ribas, I., and Delfosse, X. (2007) Habitable planets around the star Gliese 581? *Astronomy and Astrophysics*, 476, 1373-1387.

Smith, B. A., and Terrile, R. J. (1984) A circumstellar disk around Beta Pictoris. *Science*, 226, 1421-1424.

Srinivasan, G., Goswami, J. N., and Bhandari, N. (1999) $^{26}$Al in Eucrite Piplia Kalan: Plausible Heat Source and Formation Chronology. *Science*, 284, 1348-1350.

Stephan, T.(2008), Assessing the elemental composition of comet 81P/Wild 2 by analyzing dust collected by Stardust. *Space Science Reviews*, 138, 247-258.

Swindle, T. D., and Kring, D. A. (1997) Implications of small comets for the noble gas inventories of Earth and Mars. *Geophysical Research Letters*, 24, 3113-3116.

Throop, H. B., Bally, J., Esposito, L. W., and McCaughrean, M. J. (2001) Evidence for Dust Grain Growth in Young Circumstellar Disks. *Science*, 292, 1686-1698.

Tomeoka, K., and Buseck, P. R. (1990) Phyllosilicates in the Mokoia CV carbonaceous chondrite - Evidence for aqueous alteration in an oxidizing environment. *Geochimica et Cosmochimica Acta*, 54, 1745-1754.

Udry, S., Bonfils, X., Delfosse, X., Forveille, T., Mayor, M., Perrier, C., Bouchy, F., Lovis, C., Pepe, F., Queloz, D., and Bertaux, J.L. (2007) The HARPS search for southern extra-solar planets. XI. Super-Earths (5 and 8 $M_\odot$) in a 3-planet system. *Astronomy and Astrophysics*, 469, L43-L47.